\documentclass{elsart}

\usepackage{graphicx}
\usepackage{amsmath}
\usepackage{amssymb}
\usepackage{amsbsy}

\begin{document}

\begin{frontmatter}

\title{A semi--implicit Hall-MHD solver using whistler wave preconditioning}
\author{Lukas Arnold, J\"urgen Dreher and Rainer Grauer}
\address{Theoretische Physik I, Ruhr-Universit\"at Bochum, Germany}

\begin{abstract}
  The dispersive character of the Hall-MHD solutions, in particular
  the whistler waves, is a strong restriction to numerical treatments
  of this system. Numerical stability demands a time step dependence
  of the form $\Delta t\propto (\Delta x)^2$ for explicit
  calculations.  A new semi--implicit scheme for integrating the
  induction equation is proposed and applied to a reconnection
  problem. It it based on a fix point iteration with a physically
  motivated preconditioning. Due to its convergence properties, short
  wavelengths converge faster than long ones, thus it can be used as a
  smoother in a nonlinear multigrid method.
\end{abstract}

\end{frontmatter}

\section{Introduction}
In many space-, astrophysical and high temperature plasma systems
collisions do not play the most important role in describing the
departure from the ideal magnetohydrodynamics (MHD)
\begin{align}
\partial_t\rho &= -\nabla\cdot(\rho\vec{v}) \\
\partial_t\vec{v} &= -\left(\vec{v}\cdot\nabla\right)\vec{v} + \frac{\vec{j}\times\vec{B}}{\rho} - \frac{\nabla p}{\rho} \\
\partial_t\vec{B} &= -\nabla\times\vec{E}\\
\vec{j} &= \nabla\times\vec{B} \; ,
\end{align}
where $\rho$, $\vec{v}$, $\vec{B}$ and $p$ denote mass density,
velocity, magnetic field and pressure, respectively.  Typical examples
include filamentation and singularity formation, collisionless
reconnection and collisionless shocks
\cite{dreher-laveder-etal:2005,dreher-ruban-etal:2005,lottermoser-scholer:1997,shay-drake:1998,buechner-kuska:1999,horiuchi-sato:1999,birn-drake-etal:2001,ricci-lapenta-etal:2002,schmitz-grauer:2005}.
Therefore, on scales smaller than the ion inertia length additional
processes have to be taken into account in a generalized Ohm's law
\begin{equation}
  \vec{E} = \eta\vec{j} - \vec{v}\times\vec{B} +
  \frac{m_i}{Ze\rho}\left(\vec{j}\times\vec{B}-\nabla p_e\right) \;.
\end{equation}
Numerically, the most difficult term is the Hall-term:
\begin{equation}
  \vec{E}_{Hall}=\frac{m_i}{Ze\rho}\vec{j}\times\vec{B} = \frac{d_i}{\rho}\vec{j}\times\vec{B} \;.
\end{equation}
It allows for whistler wave solutions with a quadratic dispersion
relation and thus poses a severe time step restriction for a temporal
explicit discretisation.

To introduce our treatment of the Hall-term, we simplify our system
and use only this electric field in the induction equation which
decouples it from the other part of the MHD equations and yields the
following nonlinear equation
\begin{equation}
  \label{eqn:red-ind}
  \partial_t\vec{B}=-\nabla\times\left( \frac{d_i}{\rho}
    \left( \nabla\times\vec{B}\right) \times \vec{B} \right).
\end{equation}
Solutions of the linearized equations are the whistler waves mentioned
above which satisfy the dispersion relation
$\omega=\frac{d_i|\vec{B}|}{\rho}k^2$, for a constant density $\rho$ and a
guiding field magnitude $|\vec{B}|$.  Numerical approaches using explicit
schemes applied to this equation must ensure that the chosen time step
fulfills $\Delta t\propto (\Delta x)^2$, due to the
Courant-Levy-Friedrichs criterion -- $\Delta x$ denoting the grid spacing.
The CFL number is given by the ratio of the phase velocity to the grid
velocity $\left(\frac{\Delta x}{\Delta t}\right)$
\begin{displaymath}
  \mbox{CFL} =
  \frac{\omega(k)}{k}\frac{\Delta t}{\Delta x} = \frac{d_i\Delta t
    \left|\vec{B}\right|}{\rho\Delta x^2} \quad
  \Longrightarrow \Delta t = \mbox{CFL}\frac{\rho}{d_i\left|\vec{B}\right|}\Delta x^2 \; ,
\end{displaymath}
where $k=k_{max}=\frac{2\pi}{\Delta x}$ is the maximum wave number.
Thus resolving small structures, e.g. the reconnection zone, results
in large computation times, due to the unavoidable small time steps.

Implicit schemes allow to avoid this restrictive condition by
providing unconditional numerical stability. Much progress on implicit
solvers has been done by Harned and Miki\'c \cite{harned-mikic:1989}
and Chac\'on and Knoll \cite{chacon-knoll:2003}. However, the approach
of \cite{harned-mikic:1989} requires a guiding magnetic field and the
approach of \cite{chacon-knoll:2003} can't easily be adopted for
simulations with adaptive mesh refinements
\cite{berger-collela:1989,fryxell-olsen-etal:2000,dreher-grauer:2005,teyssier-fromang-etal:2006},
although work in this direction is in progress.

Here we present a simple physics based semi--implicit Crank-Nicolson
type scheme which due to its locality properties is suitable for
parallel computations as well as for use in adaptive mesh refinement
simulations. This physics based solver uses a whistler wave
decomposition to accelerate the fix-point iteration. Due to its
convergence properties it can act as a smoother for a nonlinear
multigrid scheme.

The first part of this paper presents the general numerical method
which then is specialized to one dimension. This allows us to show
analytically its convergence.  After that the nonlinear
two-dimensional case and its convergence are presented, while in the
last section our method is used to solve a two-dimensional
reconnection problem.

\section{Numerical Method}
The Richardson iteration \cite{kelley:1995} is the base of our solver. A
Richardson iteration is the most general fix point iteration for a
nonlinear equation $\vec{F}(\vec{x})=0$
\begin{equation}
  \label{eqn:richardson}
  \vec{x}^{k+1}=\vec{K}(\vec{x}^k)\quad \mbox{with} \quad \vec{K}(\vec{x})=\vec{x}-\alpha\vec{F}(\vec{x}) \;,
\end{equation}
where $k$ is the iteration index. Given a contractive map $\vec{K}$,
the $\vec{x}^k$ converge in the limit $k\rightarrow\infty$.  The rate
of convergence will in general depend on $\alpha$. The main task is to
find a suitable preconditioner adapted to the Hall-term. This can be
realized as a matrix ${\bf{P}}$
\begin{equation}
  \label{eqn:precond}
  \vec{K}(\vec{x})=\vec{x}-\alpha {\bf{P}} \vec{F}(\vec{x}) \;.
\end{equation}
In the special case of the Newton iteration ${\bf{P}}$ is the inverse of
the Jacobi matrix of $\vec{F}$. Here, we try to find a physics based
preconditioner which is more local and thus suitable for parallel and
block-adaptive calculations.

For the Crank-Nicolson type discretisation, we obtain
\begin{equation}
  \frac{\vec{B}^{n+1} - \vec{B}^n}{\Delta t} = 
  -\nabla\times\left( \frac{d_i}{\rho}
    \left( \nabla\times\vec{B}^*\right) \times \vec{B}^* \right)
\end{equation}
with $\vec{B}^* = \frac{1}{2} (\vec{B}^{n+1}+\vec{B}^n)$ and where
$\vec{B}^n$ is the magnetic field taken at the time step $n$ (time
steps are indicated by the first upper index).  The equation to be
solved reads now
\begin{align}
  \label{eqn:generalF}
  \vec{F}(\vec{B}^{n+1}) &= \frac{\vec{B}^{n+1} - \vec{B}^{n}}{\Delta t} 
  +\nabla\times\left( \frac{d_i}{\rho}
    \left( \nabla\times\vec{B}^*\right) \times \vec{B}^* \right) \\
  & = 0 \; .
\end{align}
Its solution with a given $\vec{B}^{n}$ is the magnetic field at the
next time step $n+1$.  To determine a solution we iterate equation
(\ref{eqn:generalF}) following the method given by
(\ref{eqn:richardson}). At this point we introduce an additional upper
index which defines the iteration step. So that $\vec{B}^{n+1,k}$ is
$k$-th iteration of the magnetic field for the time step $n+1$.

As mentioned above, the important point in this iteration is the
preconditioning. To motivate our preconditioner, we start with the
one-dimensional version of eqn. (\ref{eqn:red-ind}), where $\vec{B}$
depends only on $x$. In the one-dimensional case, we have the special
situation that the fix-point problem reduces to a linear one.  In this
case, our physics based preconditioner reduces to the standard
$\omega$-Jacobi iteration. In more than one dimensions the situation
is genuinely nonlinear but the smoothing properties of our
preconditioner are still similar to that of a Jacobi iteration for
linear problems.

\section{Linear 1D Case} 
Considering only the $x$ direction results in the following system of
equations
\begin{equation}
  \label{eqn:1D-anal}
  \partial_t
  \left(\begin{array}{c}
      B_{x} \\
      B_{y} \\
      B_{z} 
    \end{array}\right)
  = 
  \left(\begin{array}{c}
      0 \\
      \frac{B_0d_i}{\rho}\partial_{xx}(B_{z})\\
      -\frac{B_0d_i}{\rho}\partial_{xx}(B_{y})
    \end{array}\right)
\end{equation}
The $x$ component of $\vec{B}$ is initially set to a constant $B_0$ in
space and stays constant. The discretised equations for $\vec{F}$ are
needed for the iteration. Following equation (\ref{eqn:generalF}) and
again using
$\vec{B}^*=\frac{1}{2}\left(\vec{B}^n+\vec{B}^{n+1}\right)$ these are
given for each grid point $i$
\begin{align}
\label{eqn:1D-disc}
F_x(\vec{B}_{i}^{n+1}) =& B_{x,i}^{n+1} - B_{x,i}^{n} \nonumber\\
F_y(\vec{B}_{i}^{n+1}) =& B_{y,i}^{n+1} - B_{y,i}^{n} - c_x(B_{z,i-1}^* -2B_{z,i}^* +B_{z,i+1}^*) \\
F_z(\vec{B}_{i}^{n+1}) =& B_{z,i}^{n+1} - B_{z,i}^{n} + c_x(B_{y,i-1}^* -2B_{y,i}^* +B_{y,i+1}^*) \nonumber
\end{align}
with $c_x = \frac{B_0d_i \Delta t}{\rho(\Delta x)^2}$. To calculate
the next time step $\vec{B}_{i}^{n+1}$ we consider the iteration such that
$\lim_{k\rightarrow\infty}\vec{B}^{n+1,k}=\vec{B}^{n+1}$.

The usual choice for a preconditioning matrix would be the inverse of
the full Jacobian matrix.  Even in the one-dimensional case this would
lead to an inversion of a $2N\times2N$ matrix ($N$ is the number of
grid points). In our treatment of the Hall-term, we introduce a local
approximation to the Jacobian matrix.  The Richardson iteration
(\ref{eqn:richardson}) for $\vec{F} = 0$ reads
\begin{align}
\label{eqn:1D-disc-iter}
F_x(\vec{B}_{i}^{n+1,k}) =& B_{x,i}^{n+1,k} + r_{x,i} \nonumber\\
F_y(\vec{B}_{i}^{n+1,k}) =& B_{y,i}^{n+1,k} - \frac{c_x}{2}(B_{z,i-1}^{n+1,k} 
-2B_{z,i}^{n+1,k} +B_{z,i+1}^{n+1,k}) + r_{y,i}\nonumber \\
F_z(\vec{B}_{i}^{n+1,k}) =& B_{z,i}^{n+1,k} + \frac{c_x}{2}(B_{y,i-1}^{n+1,k} 
-2B_{y,i}^{n+1,k} +B_{y,i+1}^{n+1,k}) + r_{z,i} \; ,
\end{align}
where $r_{x,i}, r_{y,i}, r_{z,i}$ are constants depending only on the values $\vec{B}^n$.

The matrix elements of the local preconditioner are calculated by
differentiating Eqs.~(\ref{eqn:1D-disc-iter}) with respect to
$\vec{B}_{i}^{n+1,k}$; this is a differentiation with respect to the
value of $\vec{B}$ at one grid point, e.g. the derivative of
$B_{y,i+1}^{n+1,k}$ with respect to $B_{y,i}^{n+1,k}$ vanishes. This
leads to the following local preconditioner
\begin{equation}
{\bf{J}}^*_i =\frac{\partial \vec{F}(\vec{B}_i^{n+1,k})}{\partial\vec{B}_i^{n+1,k}}= \left(\begin{array}{ccc}
1 & 0 & 0 \\
0 & 1 & c_x \\
0 & -c_x & 1 
\end{array}\right).
\end{equation}
The preconditioning matrix ${\bf{J}}^*$ is now a block diagonal matrix
containing only the ${\bf{J}}_i^*$. This local construction allows an
easy inversion, which is again a block matrix and thus
$({\bf{J}}^*)^{-1}$ is local. A fast inversion of the general Jacobian
is not easy and results in a non local matrix. Using
$({\bf{J}}^*)^{-1}$ as a preconditioner results in the following local
iteration for each grid point $i$.
\begin{equation}
\vec{B}_i^{n+1,k+1} = \vec{B}_i^{n+1,k} -\frac{\alpha}{1+c_x^2}
\left(\begin{array}{ccc}
1+c_x^2 &  0& 0 \\
0 & 1 & -c_x \\
0 & c_x & 1 
\end{array}\right)
\vec{F}(\vec{B}^{n+1,k}_i) \;.
\end{equation}
This iteration is the $\omega$-Jacobi iteration, which in this case is
known to converge. This way we have created a well known iteration
scheme, but it was motivated by the physical properties of a whistler
wave. While the Jacobi iteration cannot be used for nonlinear
problems, we can transfer our iteration to the nonlinear
two-dimensional case using the same strategy. The iteration obtained
so far has the desirable property that high wavenumbers converge fast
and thus can be used as a smoother in a multigrid scheme. We
anticipate that this property translates to the two-dimensional case.

\section{Nonlinear 2D case}
\begin{figure}[ht!]
  \begin{center}
    \includegraphics[width=0.9\columnwidth]{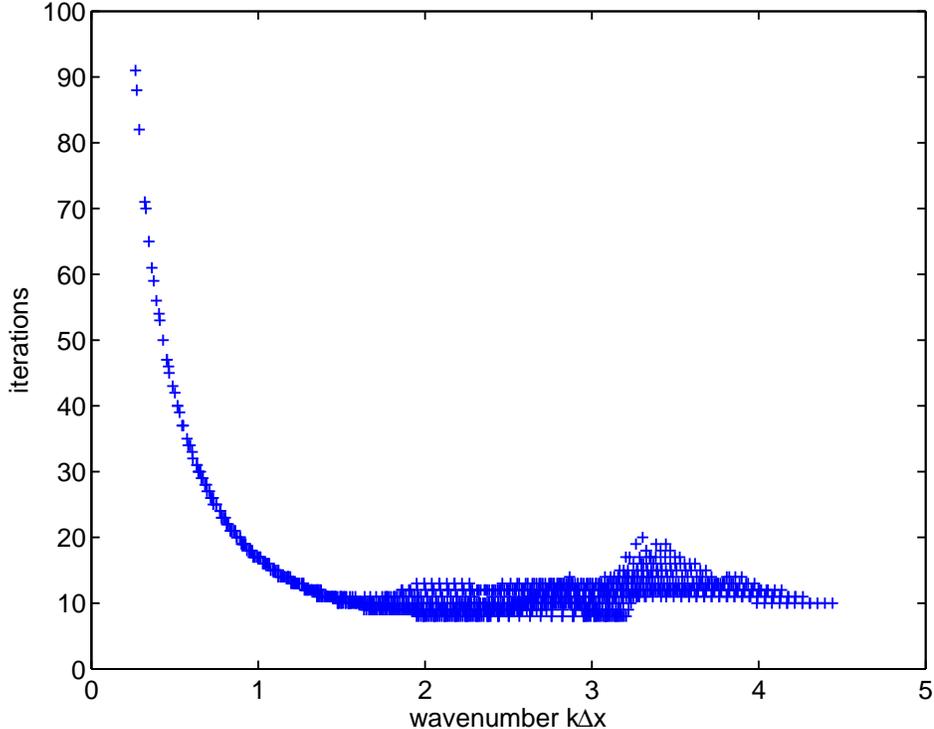}
  \end{center}
  \caption{\label{fig:iter2D}The number of iterations needed to reduce the error 
by two orders of magnitude as a function of the norm of the wave vector $k=\sqrt{k_x^2+k_y^2}$.}
\end{figure}

The strategy here will be the same as above:
\begin{equation}
\vec{B}^{n+1,k+1}_{i,j} = \vec{B}^{n+1,k}_{i,j} - \alpha {\bf{J}}_{i,j}^{*-1}\vec{F}_{i,j}(\vec{B}^{n+1,k}) \; .
\end{equation}
The local preconditioner is motivated by the one-dimensional
calculation, taking into account the two directions $x$ and $y$ of
whistler wave propagation. In two dimensions the function $\vec{F}$
can be derived the same way as the in the one-dimensional case and
takes the form
\begin{align}
  F_x &=  B_x^{n+1} - B_x^n + \frac{d_i\Delta t}{\rho}\left( {B^*_y}\partial_{yy} {B^*_z} 
    + \partial_y {B^*_y}\partial_y {B^*_z} + \partial_y {B^*_x} \partial_x {B^*_z} + {B^*_x}\partial_{xy} {B^*_z}\right) \nonumber \\
  F_y &=  B_y^{n+1} - B_y^n -\frac{d_i\Delta t}{\rho}\left( {B^*_y} \partial_{xy}{ B^*_z} 
    + \partial_x {B^*_y} \partial_y {B^*_z} + \partial_x {B^*_x} \partial_x {B^*_z} + {B^*_x} \partial_{xx} {B^*_z}\right)\nonumber \\
  F_z &=  B_z^{n+1} - B_z^n + \frac{d_i\Delta t}{\rho}\Big(\left(\partial_x {B^*_x} 
    + \partial_y {B^*_y}\right)\left(\partial_x {B^*_y} - \partial_y {B^*_x}\right) \label{eqn:2df} \\
  & \phantom{= B_z^k - B_z^n + \frac{d_i\Delta t}{\rho}\Big( (}+ {B^*_x}\left(\partial_{xx} {B^*_y} - \partial_{xy} {B^*_x}\right) 
+{B^*_y}\left(\partial_{xy} {B^*_y} - \partial_{yy} {B^*_x}\right)\Big).\nonumber
&&
\end{align}
To obtain the local preconditioner, Eq.~(\ref{eqn:2df}) has to be
discretised in space and derived with respect to $\vec{B}_{ij}^{n+1}$ which yields
\begin{equation}
{\bf{J}}_{i,j}^* = \left(\begin{array}{ccc}
1 & 0 & -c_{y,i,j} \\
0 & 1 & c_{x,i,j} \\
c_{y,i,j} & -c_{x,i,j} & 1 
\end{array}\right)
\end{equation}
with $c_{x,i,j} = \frac{d_iB^n_{x,i,j}\Delta t}{\rho{\Delta x}^2}$ and
$c_{y,i,j} = \frac{d_iB^n_{y,i,j}\Delta t}{\rho{\Delta y}^2}$.  Note
that the elements of this Jacobian matrix are not constant anymore,
but depend on $B^n_{x,i,j}$ and $B^n_{y,i,j}$. Thus again the
preconditioning is it's inverse
\begin{equation}
{\bf{J}}_{i,j}^{*-1} = \frac{1}{1+c_{x,i,j}^2 + c_{y,i,j}^2}\left(\begin{array}{ccc}
1+c_{x,i,j}^2 &  c_{x,i,j}c_{y,i,j}& c_{y,i,j} \\
c_{x,i,j}c_{y,i,j} & 1+c_{y,i,j}^2 & -c_{x,i,j} \\
c_{y,i,j} & c_{x,i,j} & 1 
\end{array}\right).
\end{equation}
To show numerically the convergence properties of this iteration, a
simple numerical experiment is used. The two-dimensional computational
domain is a periodic box in $x$- and $y$-direction. The initial
condition is a single whistler wave with a wave number in $x$- and
$y$-direction. This is done for all possible modes (combinations of
$k_x$ and $k_y$) on a $64\times64$ mesh.  The number of iterations
needed for a prescribed accuracy is plotted in Fig.~\ref{fig:iter2D}
as a function of $k=\sqrt{k_x^2+k_y^2}$.
As we already anticipated the convergence rate for the long and the
short wavelengths show the same behavior as in the one-dimensional
case.

To accelerate the convergence of the long wavelengths, this iteration
is applied as a smoothing function for the nonlinear multigrid scheme
\cite{briggs:2000}.  Using V-cycles and two pre-- and post--smoothings
in the multigrid scheme one achieves convergence already after two to
three cycles.

\section{GEM Reconnection} 
\begin{figure}[ht!]
  \begin{center}
    \includegraphics[width=0.9\columnwidth]{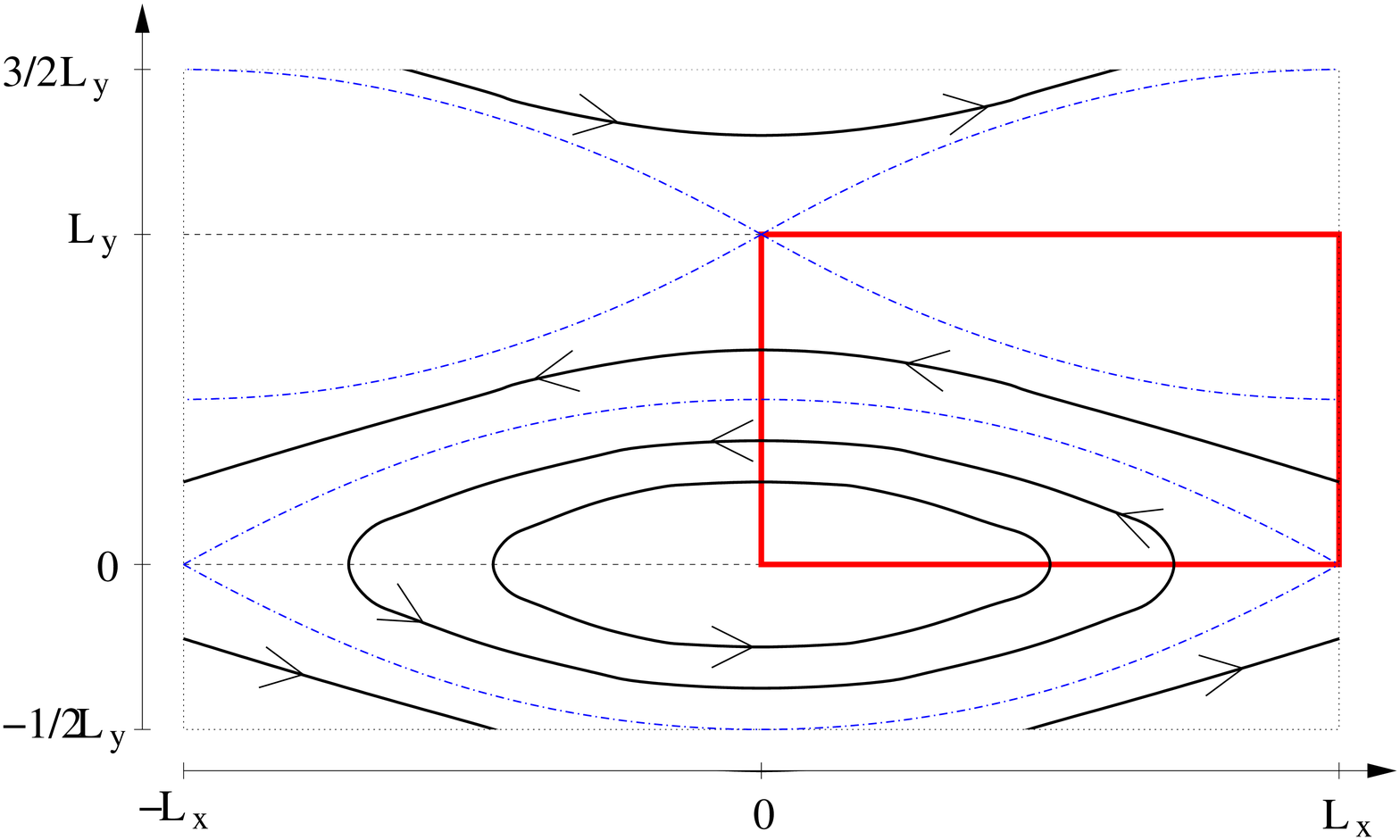}
  \end{center}
  \caption{\label{fig:double-HS}The GEM reconnection setup. The black lines 
    indicate the magnetic field, the blue
  ones the seperatrices and the red box is our computational domain.}
\end{figure}

To verify the new iteration scheme, we choose a standard reconnection
problem. A similar setup is used as in GEM reconnection challenge
\cite{birn-drake-etal:2001}. It is based on a perturbed Harris sheet.
A schematic plot and the computational domain are shown in
Fig.~\ref{fig:double-HS}.
The numerical parameter are chosen to: $L_x = 2L_y=5 d_i$ and $N_x =
2N_y=256$.  Symmetric or antisymmetric boundary conditions are applied
for all quantities and the initial conditions, including the
perturbation, are equal to the one in the GEM reconnection challenge
\cite{birn-drake-etal:2001}.

We solve the full Hall-MHD equations explicitly except for the Hall
part of the induction equation. This splits the induction equation
into the resistive MHD part (\ref{eqn:GEM-MHD}) and the part including
only the Hall term (\ref{eqn:GEM-Hall})
\begin{align}
  \label{eqn:GEM-MHD}\partial_t\vec{B}_{\footnotesize\mbox{MHD}} &= 
  \nabla\times\left(\vec{v}\times\vec{B}\right) + \eta\Delta\vec{B}\\
  \label{eqn:GEM-Hall}\partial_t\vec{B}_{\footnotesize\mbox{Hall}} &= 
  - \nabla\times\left(\frac{d_i\vec{j}\times\vec{B}}{\rho}\right)\\
  \partial_t\vec{B} &= \partial_t\vec{B}_{\footnotesize\mbox{MHD}} + \partial_t\vec{B}_{\footnotesize\mbox{Hall}} \; .
\end{align}
The time stepping was performed with a standard second order
Runge-Kutta method.

The maximum time step for the semi--implicit simulations is given by
the linear Alfv\'en wave dispersion relation $\omega = v_A k$. There
is no possibility to use larger time steps than these, because the
Alfv\'en waves must be well resolved.

To compare the results obtained with the fully explicit simulation and
with our semi--implicit treatment of the Hall term, we choose as a
physically relevant measure the reconnected flux $\psi$ given in our
setup by
\begin{equation}
  \psi = \int^{L_x}_0B_y\ dx.
\end{equation}
Four different simulation runs have been done, an explicit one,
used as the reference run,
and three semi--implicit ones.
The time steps and corresponding CFL numbers are shown in Table \ref{tab:dt}.
\begin{table}[h]
  \begin{center}
    \begin{tabular}{l|c|c|c}
      & $\Delta t\cdot10^{-3}$ & CFL number & $\frac{\Delta t_{imp}}{\Delta t_{exp}}$\\
      \hline 
      explicit & 0.2 & 0.2 & 1 \\
      semi--implicit & 2.0 & 2 & 10 \\
      semi--implicit & 4.0 & 4 & 20 \\
      semi--implicit & 8.2 & 8.2 & 41
    \end{tabular}
  \end{center}
  \caption{\label{tab:dt} Time steps chosen for the simulations. The CFL number is
    based on the whistler wave dispersion relation.}
\end{table}

\begin{figure}[ht!]
  \begin{center}
    \includegraphics[width=0.9\columnwidth]{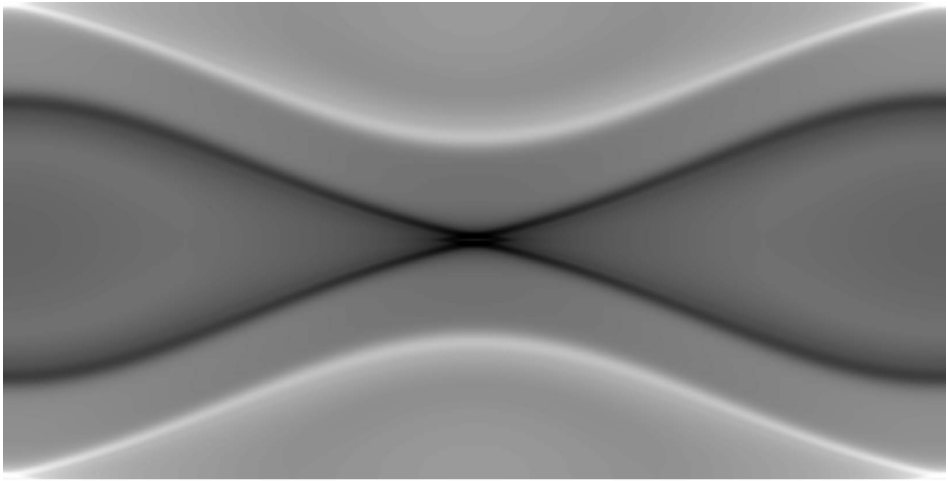}
  \end{center}
  \caption{\label{fig:jz}The $z$-component of the electric current density at $t=12$.}
\end{figure}

\begin{figure}[ht!]
  \begin{center}
    \includegraphics[width=0.9\columnwidth]{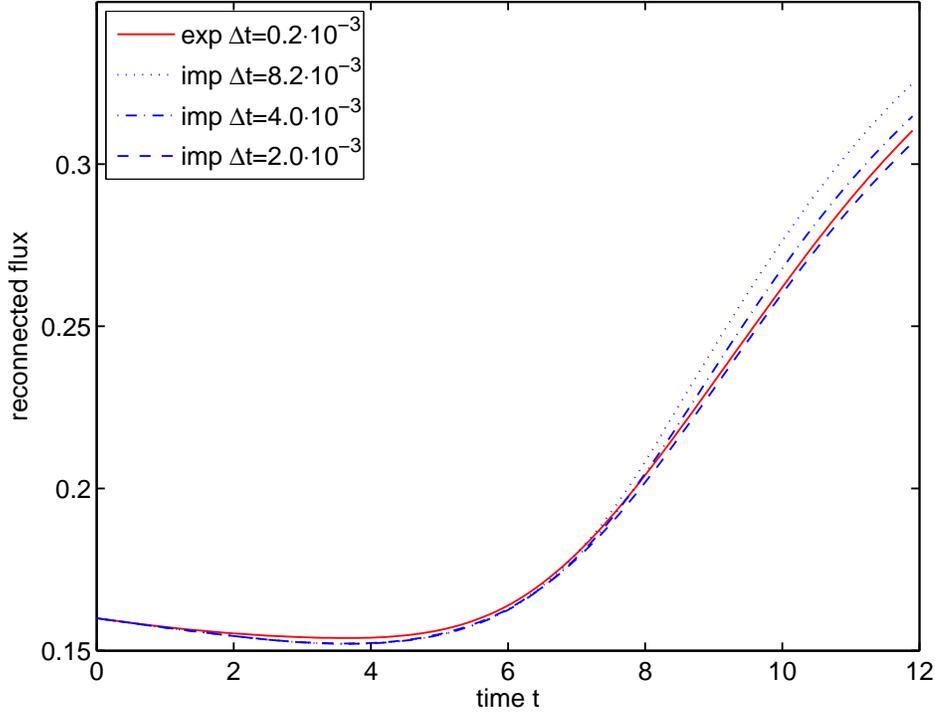}
  \end{center}
  \caption{\label{fig:recrate}The temporal evolution of the reconnected flux for different time steps and schemes.}
\end{figure}

Fig.~\ref{fig:jz} shows the developed structure of the $z$-component
of the electric current density $\vec{j} = \nabla\times\vec{B}$ at
time $t=12$ obtained from a semi--implicit simulation.
The corresponding reconnected flux (Fig.~\ref{fig:recrate}) obtained
from the explicit reference simulation and the semi--implicit runs
show that the the semi--implicit simulations result in nearly
identical reconnection rates and that, by reducing the time step, they
converge to the values obtained from the explicit simulation.

\section{Summary}
We presented a semi--implicit iterative method for solving the Hall
part of the induction equation to overcome the time step restriction
resulting from the quadratic whistler wave dispersion relation. The
method utilizes a simple precondioner based on the whistler wave
dispersion. The iteration scheme based on this precondioner has the
two desirable features of being local and possessing strong high
frequency smoothing. Therefore, this method can easily be implemented
in a nonlinear multigrid solver. Due to the locality property, it is
also best suited for adaptive and parallel simulations. The part of
execution time of Hall term turns out to be about 7 times longer than
the time needed for ideal MHD part. Using a time step 40 times larger
than necessary for an explicit treatment, this results in an 80\%
reduction of the computation time for the GEM setup achieving nearly
identical results as an expensive explicit simulation.

\ack This work benefited from support through SFB 591 of the Deutsche
Forschungsgesellschaft and the HGF virtual Institute VH-VI-123.


\end{document}